\documentclass[a4paper,conference]{IEEEtran}

\usepackage{xcolor,soul,framed} 

\colorlet{shadecolor}{yellow}
\usepackage[pdftex]{graphicx}
\graphicspath{{../pdf/}{../jpeg/}}
\DeclareGraphicsExtensions{.pdf,.jpeg,.png}

\usepackage[cmex10]{amsmath}
\usepackage{array}
\usepackage{mdwmath}
\usepackage{mdwtab}
\usepackage{eqparbox}
\usepackage{url}
\usepackage{amssymb}
\usepackage[euler]{textgreek}
\usepackage{autobreak}
\usepackage{tabularx}

\begin{document}
\bstctlcite{IEEEexample:BSTcontrol}
    \title{StegColNet: Steganalysis based on an ensemble colorspace approach}
  \author{Shreyank N Gowda \\ University of Edinburgh \\ Edinburgh EH8 9YL, UK \\ (e-mail: s1960707@ed.ac.uk). \and
  \author CChun Yuan \\ Tsinghua University \\ Shenzhen 518055, China \\ (e-mail: yuanc@sz.tsinghua.edu.cn). \ 

  \thanks{Supported by NSFC project Grant No. U1833101, Shenzhen Science and Technologies project under Grant No. JCYJ20160428182137473 and the Joint Research Center of Tencent and Tsinghua.}
  \thanks{Shreyank N Gowda is with the University of Edinburgh, Edinburgh EH8 9YL, UK (e-mail: s1960707@ed.ac.uk).}
  \thanks{Chun Yuan is with Tsinghua University, Information Science and Technology, Shenzhen 518055, China (e-mail: yuanc@sz.tsinghua.edu.cn).}%
}

\markboth{IEEE SIGNAL PROCESSING LETTERS, VOL. 27, 2020
}{Gowda \MakeLowercase{\textit{et al.}}: StegColNet}

\maketitle

\begin{abstract}
Image steganography refers to the process of hiding information inside images. Steganalysis is the process of detecting a steganographic image. We introduce a steganalysis approach that uses an ensemble color space model to obtain a weighted concatenated feature activation map. The concatenated map helps to obtain certain features explicit to each color space. We use a levy-flight grey wolf optimization strategy to reduce the number of features selected in the map. We then use these features to classify the image into one of two classes: whether the given image has secret information stored or not. Extensive experiments have been done on a large scale dataset extracted from the Bossbase dataset. Also, we show that the model can be transferred to different datasets and perform extensive experiments on a mixture of datasets. Our results show that the proposed approach outperforms the recent state of the art deep learning steganalytical approaches by 2.32 percent on average for 0.2 bits per channel (bpc) and 1.87 percent on average for 0.4 bpc.
\end{abstract}


\section{Introduction}
\label{sec:intro}

Steganography is a means of covert communication in which secret information is embedded into some form of digital media, such as an image, video or text file [3].  Usually, this form of embedding is done such that there is no
apparent perceptible change in the embedding file. In multimedia security, steganography forms a critical research topic [4]. The difference between steganography and cryptography is that in cryptography data is encrypted and although difficult to break, raises a doubt in the mind of an attacker about the presence of secret information. Steganography, on the other hand, aims to reduce the risk of being
detected.

In general, images are considered as the embedding medium due to minute changes in an image being imperceptible to the human eye [4].  There are three main
properties that a steganographic algorithm should possess:
security, robustness, and capacity. In case of an image
steganographic algorithm, security would mean how securely the algorithm can hide information, i.e., how little visual change is caused on an image using an image steganography algorithm. Robustness refers to the invariability of
the steganographic algorithm when an image is subject of
different transforms such as scaling, resizing, rotation, etc.
The capacity for a steganographic algorithm represents the amount of data that can be embedded in an image before there is a noticeable visual change in the image [5]. Steganalysis is the process of detecting if a given image has information hidden in it or not [27]. In this regard, we can convert this problem into that of a simple classification problem. To detect if an image is embedded with information we propose the use of an ensemble color space model. Recently, it was seen an ensemble colorspace model [1] obtained excellent results on large scale image classification datasets such as imagenet [2]. Based on [1] we propose a novel steganalysis approach.

 Steganalysis is the process of detecting if a given image
has information hidden in it or not. In this regard, we
can convert this problem into that of a simple classification
problem. To detect if an image is embedded with information, we propose the use of an ensemble color space model. 

We do the following:
\begin{itemize}
    \item We use a colorspace approach to determine if an image is hiding information or not. We use ColorNet [1] and take the final activation map from each colorspace. 
    \item We use weighted averaging to obtain a single feature map from all the individual feature maps that are generated by each colorspace.
It was seen [1] that each color space had features explicit to themselves and this would help us detect minute changes in the image.
\item We then use a levy-flight grey wolf optimization method (meta-heuristic approach) to select a smaller subset of features.
Using these features, we classify the given image into one of two classes: containing concealed information or not.

\end{itemize}

\section{Related Work}
\label{sec:rw}

\subsection{STEGANOGRAPHY}

Most steganography algorithms can be expressed in figure 1. An image is broken down to it’s RGB (Red Green
Blue) channels and pixels in the individual channels are
modulated with some cost function ’C’ which embeds information into that channel. The most straightforward
steganography algorithm is the LSB (Least Significant Bit)
algorithm. Here, as the name suggests the least significant
bit is taken, and one bit of information is stored (either as a
1 or a 0).

\begin{figure*}
    \begin{center}
        
    \includegraphics[width=0.9\linewidth]{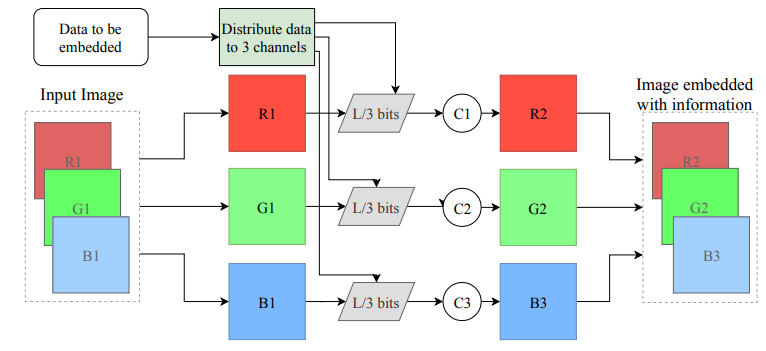}
    \caption{Pipeline of a standard steganography algorithm}
    \label{fig:my_label}
    \end{center}
\end{figure*}

Steganography algorithms can be classified broadly into four categories: 1) cover image size 2) embedding domain-based algorithms 3) nature of retrieval based algorithms 4) adaptive steganographic algorithms. In the case of 2-D images, the information is embedded onto the 2-D plane of the cover image. This embedding can be done over transform domain coefficients (such as discrete cosine transforms, Fourier transforms, etc.) or on the spatial domain (an example is LSB). The 3-D approaches essentially follow the same general procedure. However, the procedure is repeated on multiple planes (for instance RGB in a color image has 3 planes that can embed information). Image steganography on 3-D images can be made in either geometrical domain [5], representation domain [6] or topological domain [7]. 

Some of the transform-based steganographic algorithms include discrete Fourier transform (DFT) [9], discrete cosine transform (DCT), discrete wavelet transform [10], complex wavelet transform [11] among others. Here, frequency coefficients obtained after applying transforms are used to hide secret bits. Along with the security being improved, these algorithms are robust to image compression, cropping, scaling, etc. 
Off late, machine learning approaches have been proposed such as SVM (Support Vector Machine)[12], genetic algorithm approaches [13], neural network-based steganography [14]. Though these approaches are black-box approaches, they have shown good results.

\subsection{STEGANALYSIS}

Steganalysis is the method of trying to either determine a stego image (image where information is hidden) or extract the secret information. Our method deals with the former. We treat the problem at hand to be a classification problem, wherein, each image either contains some hidden information or not. 

There are two basic approaches to steganalysis: signature steganalysis and statistical steganalysis. Signature steganalysis is the method wherein patterns, or signatures relevant to various steganographic algorithms are searched for. The presence of a pattern indicating that secret information
is being hidden in the image. The quintessential process
here is the repetition of patterns due to embedded secret information. The statistical approach searches for mathematical results to determine if the information is being hidden.

Signature steganalysis is further classified into specific embedding [16] and universal blind steganalysis [15]. Specific embedding approaches are impractical because we need to know what steganography approach has been used to embed information. Hence, universal blind steganalysis [8,17] is preferred. These approaches help in the extraction of high dimensional features. 
However, the curse of dimensionality occurs. Hence, a need to reduce feature size occurs. Some commonly used algorithms to do the same include wrappers, filters, etc. Filters are less complex; however, they perform poorly. Wrapper methods evaluate feature subset using predictive models [18]. However, wrappers are complex and time-consuming. 

To overcome this, meta-heuristic approaches have been deployed. These approaches solve optimization problems by utilizing natural phenomena [19-20]. It was seen that Grey Wolf Optimization (GWO) performed better than other metaheuristic approaches for solving non-linear problems in a multi-dimensional space [19]. However, it has a slow convergence rate and gets trapped in local optima at times. It has been seen that GWO can be optimized by modifying it's parameter \textbf{A} to obtain a quick convergence rate, better convergence precision and higher agility for global searching.

\section{Proposed Approach}

\subsection{Overall architecture and effect of using color spaces}

We consider steganalysis as a 2 class classification problem. The overall architecture is described in figure 2. The experimental analysis along with details regarding training set etc are explained in the next section. Recently, the effect of color spaces on image classification has been explored [1]. It was seen that individual color spaces inherited classification features explicitly to themselves. This helped us ponder about the ability to extract information in an image where there is secret information being embedded. Colornet [1] being an ensemble model, that could extract features specific to each colorspace, was an excellent choice to utilize to help us in determining if an image could have information hidden in it. The output of Colornet is a high-dimensional vector, which causes a computationally intensive execution. To reduce the number of features selected we have to use an optimization approach for feature selection. Figure 1 shows the architecture of the model.

\begin{figure*}[t]
\begin{center}
   \includegraphics[width=0.9\linewidth]{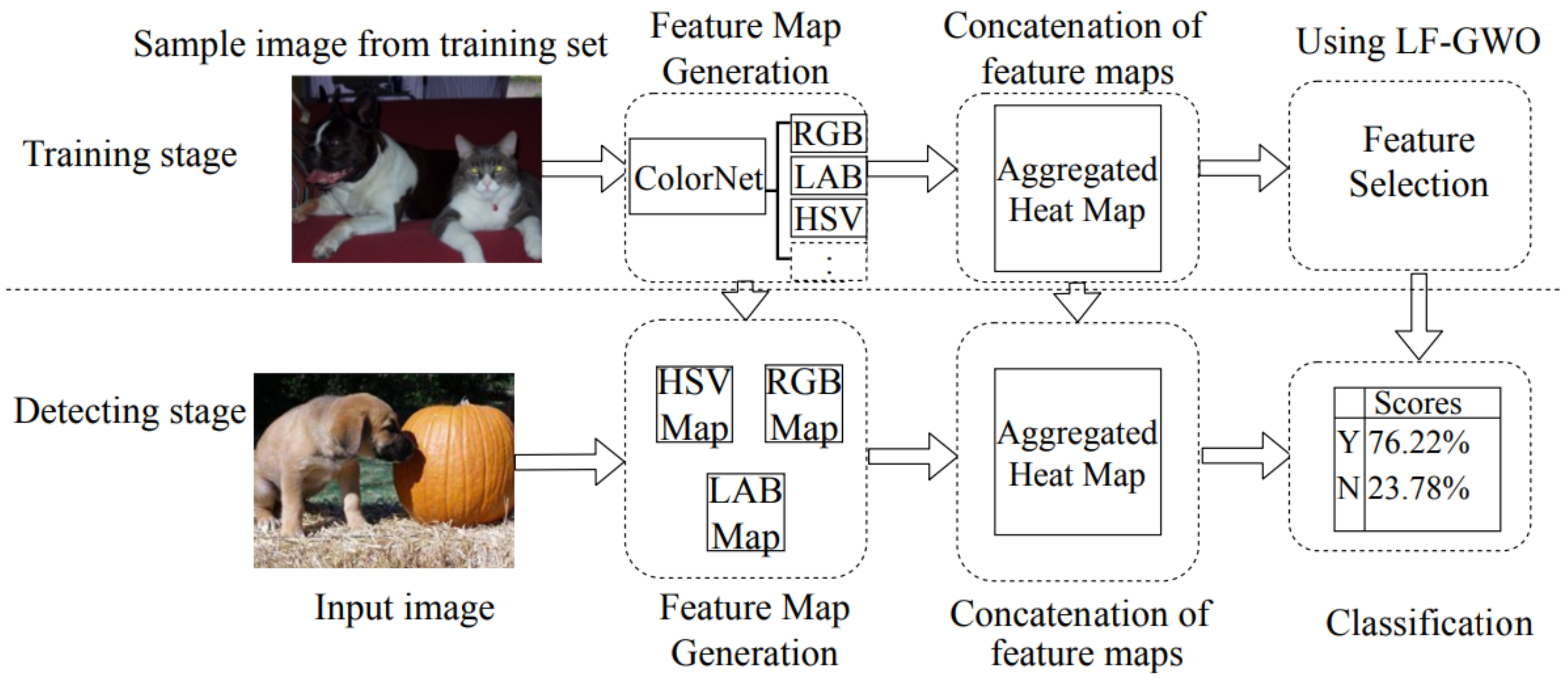}
\end{center}
   \caption{Two phases involved in the overall architecture of the model: training the model using colornet and detecting stego-image using feature map aggregation}
\label{fig:long}
\label{fig:onecol}
\end{figure*}

\subsection{Optimization process for feature selection}

\subsubsection{Feature selection using LF-Grey Wolf optimization} 

In GWO, the head of the pack is the \textalpha. The next level of the hierarchy is \textbeta, \textdelta \ and finally followed by \textomega. GWO models the social hierarchy and mathematically illustrates the hunting procedure as an optimization problem. 
If \textbf{X}$_{\text{p}}$(t) and \textbf{X}(t) represent the position of prey and wolf at iteration 't', we can mathematically model the encircling process [19] with two coefficients \textbf{A} and \textbf{C} as shown in (1). \textbf{A} and \textbf{C} are calculated by (2).

\setlength{\belowdisplayskip}{0pt} \setlength{\belowdisplayshortskip}{0pt}
\setlength{\abovedisplayskip}{0pt} \setlength{\abovedisplayshortskip}{0pt}

\begin{equation}
\mathbf{D} = |\mathbf{C}.\mathbf{X}_{p}(t)-\mathbf{X}(t)| ; \mathbf{X}(t+1)=\mathbf{X}_{p}(t)-\mathbf{A.D}
\end{equation}

\begin{equation}
\mathbf{A = 2a.r_{1}-a} ; \mathbf{C=2.r_{2}}
\end{equation}

Here, \textbf{r}$_{\text{1}}$ and  \textbf{r}$_{\text{2}}$ are random vectors in [0,1],  \textbf{a} is a parameter that decreases linearly from 2 to 0 over iterations and also helps to control step size \textbf{D} of a grey wolf. 
 Implementation of the end of the hunting process is done by decreasing the value of \textbf{A} which in turn depends on \textbf{a}. Once \textbf{a} turns zero, it means that the wolves have stopped moving. The linear decrease in \textbf{A} helps to exploit search space with minimal exploration. Hence, this traps a local optimum.

The size of the aggregated feature map creates an issue in terms of the complexity of the algorithm and the overall time needed for execution. To deal with this, we propose the use of levy flight-based grey wolf optimization (LF-GWO) for feature selection based on Levy probability function in (3). Here, \textmu \ represents position parameter, \textgamma \ represents scale parameter and \texteta \ represents the collection of samples in the distribution. The above equation holds good for all positive values of \textmu \ and 0 otherwise.

\begin{equation}
\mathbf{L(\eta ,\gamma ,\mu )=\frac{\sqrt{\gamma }}{2\pi }exp[-\frac{\gamma }{2(\eta -\mu )}]\frac{1}{(\eta -\mu )^{\frac{3}{2}}}}
\end{equation}

The parameter \textbf{A} is modified by the Levy flight function as \textbf{A = L(S)*r1}. This makes \textbf{A} take up values in a non-linear decrease. \textbf{S} is the position of the wolf and \textbf{r1} is a random vector.

\subsubsection{Choice of optimization function}

The reason for selection of LF-GWO is based in the statistical results obtained in [21]. It was seen that for 15 defined benchmark functions, the wilcoxon rank sum test of LF-GWO outperforms existing optimization approaches in terms of mean fitness values.

Figure 3 represents a comparison of the LF-GWO with Grey Wolf Optimization (GWO), Gravitational search algorithm (GSA), particle swarm optimization (PSO) and  fast evolutionary programing (FEP) using a boxplot and a graph showing how quickly the convergence of the best fitness value is obtained with respect to the number of iterations. The box plot represents the benchmark function defined in equation 4 and the convergence map that of the function defined in equation 5.

\begin{equation}
    F_{1}(X)=\sum_{i=1}^{d}x_{i}^{2}
\end{equation}
\begin{multline}
    F_{9}(X)= sin^{2}(3\pi x_{1}) \sum_{i=1}^{d}(x_{i}-1)^{2}\\
    \left [1+sin^{2}(3\pi x_{i}+1)  \right ]+(x_{d}-1)^{2}\left [1+sin^{2}(2\pi x_{d})  \right ]\\ +\sum_{i=1}^{d}u(x_{i},5,100,4)
\end{multline}
\begin{figure*}[t]
\begin{center}
   \includegraphics[width=0.9\linewidth]{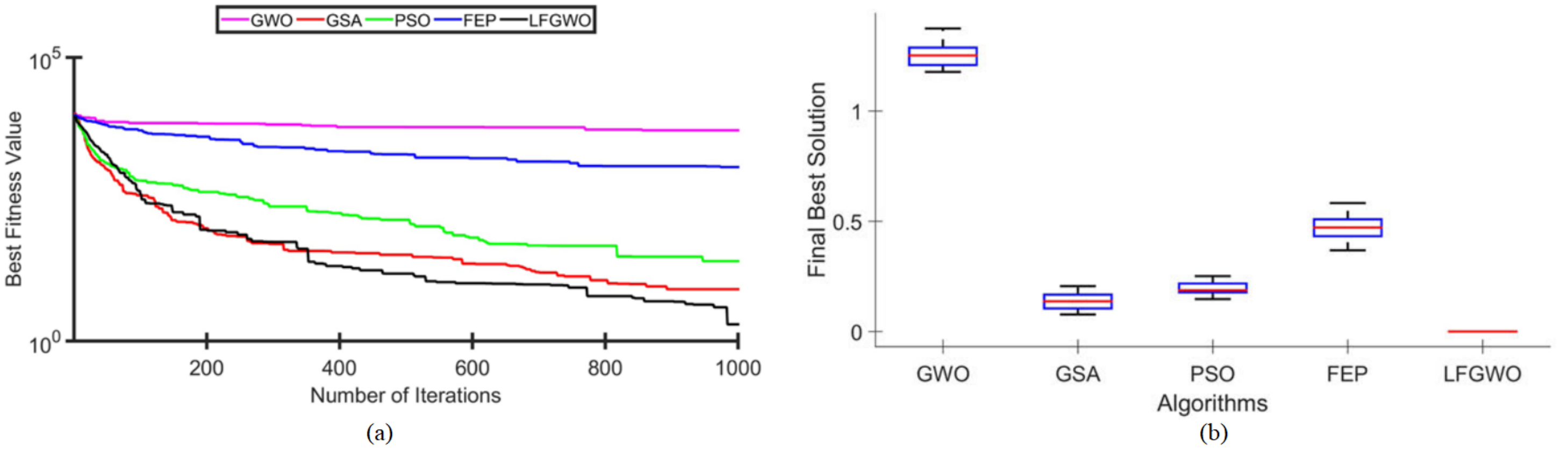}
\end{center}
   \caption{(a) Maps the convergence of the best fitness value with respect to number of iterations (b) Shows the box plot for the final best solution. Taken from [21]. Both graphs are representative of one benchmark function.}
\label{fig:long}
\label{fig:onecol}
\end{figure*}

\section{Experimental Analysis}

\subsection{Datasets and training}

Most commonly used steganalysis datasets are the Bossbase [22] and BOWS2 [23]. Each contains 10000 grayscale images. However, the approach proposed is dependent on color, and as such, we use a dataset with color images. Hence, starting with the 10000 images of Bossbase [22] dataset, we generate a dataset by following the process done in [24]. We downsampled the full-resolution images to a size of 512x512. We then followed the process in [25], so that the training and testing scenarios were conducted in a similar environment. In [25], two datasets were created by using two demosaicing algorithms: Patterned pixel grouping (PPG) and Adaptive Homogeneity-Directed (AHD) and named BOSS-PPG-LAN and BOSS-AHD-LAN correspondingly. Further, by removing the down-sampling method, we can obtain two more datasets: BOSS-PPG-CRP and BOSS-AHD-CRP. By pairing a demosaicing algorithm with bilinear or bicubic kernels, we obtain four more datasets: BOSS-PPG-BIL, BOSS-AHD-BIL, BOSS-AHD-BIL, and BOSS-AHD-BIC.

We train our model by utilizing mini-batch stochastic gradient descent with the following parameters: learning rate : 0.0001, weight decay : 0.0005, step size : 5000, momentum: 0.75, gamma : 0.75, batch size: 32, maximum iterations: 40 x 104. Testing of the trained model was done for every 5000 iterations and accuracy in 40 x 104 iterations. HILL, SUNIWARD, CMD-C-SUNIWARD and CMD-C-HILL: 4 state of the art color steganography algorithms, were used as attacking targets for experimental analysis. The embedding payload was set to 0.2 bpc (bits per channel/band pixel) and 0.4 bpc. In order to select the most challenging scenarios and also follow similar conditions for result comparison, we followed the process executed in WISERNet [25].

\subsection{Results comparison}

To compare our results, we considered three deep learning approaches for color steganalyzers, that are widely considered state of the art approaches: WISERNet [25], Deep Hierarchical Representations (DHR) [26] and Deep-CNN [27]. Experiments were conducted on the same datasets and using similar resources for a fair comparison. Popular steganography methods such as SUNIWARD [28], MiPOD [29], HILL [30] adopt an additive embedding distortion approach for minimizing framework [31]. Recently, CMD-C was proposed [32] by improvising the CMD approach for color images. We denote the CMD-C method using SUNIWARD and HILL as CMD-C-SUNIWARD and CMD-C-HILL respectively. Although DHR [26] and D-CNN [27] can be executed in channel-wise convolution, normal convolution and input concatenation as seen in [25], we show results only for the normal convolution as WiserNet [25] outperforms DHR and D-CNN in all cases. We also compare results with the Pixel Vector Cost (PVC) [33] and channel gradient correlation (CGC) [34].

The parameters used in terms of batch size and iterations were the same for all the comparisons. The other parameters were used as described in the original paper. Each experiment constituted 75 percent training images, i.e., 7500 images and 2500 images were used for testing. All experiments were performed 10 times and the average accuracy of testing was used. Table 1 compares the results of our approach with WISERNet (W-Net) [25], DHR [26], D-CNN [27], on BOSS-PPG-LAN (B-P-L), BOSS-PPG-BIC (B-P-Bc), BOSS-PPG-BIL (B-P-Bl), BOSS-AHD-BIC (B-A-Bc) and BOSS-AHD-BIL (B-A-Bl) with 0.2 bpc and table 2 with 0.4 bpc. As can be seen, the proposed method outperforms other state of the art methods for all but one case and also the percentage increase in detection is significant when patterned pixel grouping is performed on the datasets.

\begin{table}[htb]
\begin{center}
\caption{Comparison of results for CMD-C-HILL stego images with 0.2 bpc. D-CNN is executed with 30 fixed SRM kernels. The best results are represented in bold font. }
\begin{tabular}{l c c c c c c}
\hline
Dataset & DHR & D-CNN & W-Net & CGC & PVC & Proposed \\
\hline
B-P-L & 0.6474 & 0.6562 & 0.7139 & 0.7231 & 0.7120 & \textbf{0.7741}\\
B-P-Bc &  0.6589 &   0.7124 &  0.7318 & 0.7278 & 0.7657 & \textbf{0.7912}\\
B-P-Bl & 0.7611 & 0.7487 & 0.8033 & 0.8120 & 0.8068 & \textbf{0.8316}\\
B-A-Bc & 0.6614 &  0.6627 & \textbf{0.7369} & 0.7168 & 0.7211 & 0.7368 \\
B-A-Bl & 0.7622 & 0.7647 & 0.8022 & 0.7981 & 0.7764 & \textbf{0.8044} \\
\hline
\end{tabular}
\end{center}

\end{table}

\begin{table}[htb]
\begin{center}
\caption{Comparison of results for CMD-C-HILL stego images with 0.4 bpc. D-CNN is executed with 30 fixed SRM kernels. The best results are represented in bold font. }
\begin{tabular}{l c c c c c c}
\hline
Dataset & DHR & D-CNN & W-Net & CGC & PVC & Proposed \\
\hline
B-P-L &  0.7568 & 0.7941 & 0.8361 & 0.8268 & 0.8148 & \textbf{0.8724}\\
B-P-Bc &  0.7732 &   0.8068 &  0.8435 & 0.8314 & 0.8514 & \textbf{0.8814}\\
B-P-Bl & 0.87211 & 0.9045 & 0.9169 & 0.9165 & 0.9056
 & \textbf{0.9381}\\
B-A-Bc & 0.7728 &  0.8141 & 0.8448 & 0.8412 & 0.8378 & \textbf{0.8468} \\
B-A-Bl & 0.8738 & 0.9067 & \textbf{0.9144} & 0.9044 & 0.9022 & 0.9088\\
\hline
\end{tabular}
\end{center}

\end{table}

Further experimental analysis is done by mixing datasets as shown in [27]. Table 3 shows how the datasets were mixed. We further label the datasets in roman numerals for simplicity to display in the comparison of steganalyzers in table 4 and 5. BPL, BPBc, BPBl, BABc, BABl, BAL are further abbreviations of BOSS-PPG-LAN, BOSS-PPG-BIC, BOSS-PPG-BIL, BOSS-AHD-BIC, BOSS-AHD-BIL and BOSS-AHD-LAN.

\begin{table}[htb]
\begin{center}
\caption{Representation of mixture of datasets. \checkmark implies dataset has been selected and - implies otherwise.  }
\begin{tabular}{l c c c c c c}
\hline
Name & BPL & BPBc & BPBl & BABc & BABl & BAL \\
\hline
Set-I & \checkmark & \checkmark & \checkmark & - & - & - \\
Set-II &  - & - & - & \checkmark & \checkmark & \checkmark\\
Set-III &  \checkmark & - & - & - & - & \checkmark \\
Set-IV & \checkmark & \checkmark & \checkmark & \checkmark & \checkmark & \checkmark\\
\hline
\end{tabular}
\end{center}
\end{table}

Similarly to tables 1 and 2, table 4 compares results on the above-mentioned mixture of datasets with 0.2 bpc. Table 5 compares the results with 0.4 bpc. As can be seen, the proposed method outperforms recent state of the art approaches, by a significant margin.

\begin{table}[htb]
\begin{center}
\caption{Comparison of results for CMD-C-HILL stego images with 0.2 bpc on mixture of datasets. D-CNN is executed with 30 fixed SRM kernels. The best results are represented in bold font. }
\begin{tabular}{l c c c c c c}
\hline
Dataset & DHR & D-CNN & W-Net & CGC & PVC & Proposed \\
\hline
Set-I &  0.7237 & 0.7259 & 0.7675 & 0.7712 & 0.7734 & \textbf{0.8029}\\
Set-II &  0.7214 &   0.7217 &  0.7714 & 0.7710 & 0.7684 & \textbf{0.8026}\\
Set-III & 0.6722 & 0.6865 & 0.7284 & 0.7412 & 0.7388
 & \textbf{0.7648}\\
Set-IV & 0.7164 &  0.7182 & 0.7671 & 0.7782 & 0.7684 & \textbf{0.8048} \\
\hline
\end{tabular}
\end{center}

\end{table}

\begin{table}[htb]
\begin{center}
\caption{Comparison of results for CMD-C-HILL stego images with 0.4 bpc on mixture of datasets. D-CNN is executed with 30 fixed SRM kernels. The best results are represented in bold font. }
\begin{tabular}{l c c c c c c}
\hline
Dataset & DHR & D-CNN & W-Net & CGC & PVC & Proposed \\
\hline
Set-I &  0.8241 & 0.8289 & 0.8594 & 0.8788 & 0.8641 & \textbf{0.9041}\\
Set-II &  0.8231 &   0.8417 &  0.8806 & 0.8762 & 0.8661 & \textbf{0.9021}\\
Set-III & 0.7812 & 0.7892 & 0.8316 & 0.8411 & 0.8421 
 & \textbf{0.8598}\\
Set-IV & 0.8161 &  0.8214 & 0.8893 & 0.8796 & 0.8812 & \textbf{0.9013} \\
\hline
\end{tabular}
\end{center}

\end{table}

\section{Conclusion}

With recent developments of color based steganography algorithms, the need for a powerful steganalyzer is needed. We saw recently, that an ensemble model of colorspaces has a significant impact on classification results. We propose StegColNet as a powerful color image steganalyzer. We employ an ensemble colorspace strategy to determine if an image is protecting information or not.
We use ColorNet and take the final activation map from each colorspace.
We use weighted averaging to obtain a single feature map from all the feature maps that are generated by each colorspace.
We then use a levy-flight grey wolf optimization method to select a smaller subset of features.
Using these features, we classify the given image into one of two classes: containing concealed information or not.

\bibliographystyle{IEEEbib}
\bibliography{strings,refs}

\end{document}